\documentclass[nofootinbib,amsfonts,prd,aps]{revtex4}

\begin{document}

\title{A local Hamiltonian for spherically symmetric gravity coupled
  to a scalar field}

\author{ N\'estor \'Alvarez$^{1}$, Rodolfo Gambini$^{1}$,
Jorge Pullin$^{2}$}
\affiliation {
1. Instituto de F\'{\i}sica, Facultad de Ciencias, 
Igu\'a 4225, esq. Mataojo, Montevideo
, Uruguay. \\
2. Department of Physics and Astronomy, Louisiana State University,
Baton Rouge, LA 70803-4001}

\begin{abstract}
  We present a gauge fixing of gravity coupled to a scalar field in
  spherical symmetry such that the Hamiltonian is an integral over
  space of a local density. Such a formulation had proved elusive over
  the years.  As in any gauge fixing, it works for a restricted
  set of initial data. We argue that the set could be large enough to
  attempt a quantization the could include the important case of an
  evaporating black hole.
\end{abstract}

\maketitle

Spherically symmetric gravity coupled to a scalar field has been an
arena where many seminal ideas of black hole physics originated,
through classical and semi-classical treatments. The full quantization
of the model has resisted analysis, in part due to the complexity of
the Hamiltonian structure of the system. If one could gain control of
this model it would be a superb scenario to test key ideas about black
hole evaporation. The first attempt to treat the problem quantum
mechanically was carried out by Berger, Chitre, Nutku and Moncrief
\cite{bcnm} and further developed by Unruh \cite{unruh}. The resulting
Hamiltonian was intractable enough that Unruh remarked ``I present it
here in the hope that someone else may be able to do something with
it.'' More recently, Husain and Winkler and Daghigh, Kunstatter and
Gegenberg \cite{huwi}, using Painlev\'e--Gullstrand coordinates
simplified somewhat Unruh's treatment. All these efforts, however,
failed to produce a local Hamiltonian. We would like to show that
using Ashtekar's new variables a gauge fixing can be found that yields
a Hamiltonian that is the spatial integral of a Hamiltonian density.
A similar gauge fixing can be carried out in traditional variables
\cite{unruhpersonal} . It also appears to apply in other $1+1$ models,
like the Callan--Giddings--Strominger--Horowitz black holes
\cite{cghs}. We do not have a clear explanation as to why it seems to
apply in such generality, it appears to be related to the possibility
of defining a mass function \cite{kuchar,kunstatter}. 

The subject of spherical symmetry with Ashtekar's new variables has
been discussed in many instances. We will not carry out a full discussion
here. We refer the readers to the literature. This is just a minimal
introduction in order to make the paper self-consistent.
The topology of the spatial manifold will be chosen of the form
$\Sigma=R^+\times S^2$. We will use a radial coordinate $x$ and study
the theory in the range $[0,\infty]$. The case
in which there is a horizon at $x=0$ can be treated with suitable boundary
conditions. 

The formalism for dealing with spherically symmetric gravity with 
Ashtekar's new variables was discussed by Bojowald and Swiderski\cite{boswi}
and also in our recent paper\cite{exterior}.
It is best to make several changes
of variables to simplify things and improve asymptotic behaviors.
We will not go through all these steps here. It suffices to notice that at the end of the process one is left with two pair of canonical
variables $E^\varphi$ and ${K}_\varphi$ (in our recent
paper\cite{exterior} called $\bar{A}_\varphi$), and $E^x$ and $K_x$ 
that are related to the traditional canonical variables in spherical symmetry
\begin{equation}
ds^2=\Lambda^2 dx^2+R^2 d\Omega^2=\frac{\left(E^\varphi\right)^2}{E^x}
dx^2 + E^x d\Omega^2.
\end{equation}
and $P_\Lambda= \sqrt{E^x} K_\varphi/(2\gamma)$ where $\gamma$ is the
Barbero--Immirzi parameter and $P_\Lambda$ is the momentum canonically
conjugate to $\Lambda$ and we are considering the positive branch of
$E^x$.  One also has that the conjugate momentum to the variable $R$
is given by $P_R=\sqrt{E^x}\frac{K_x}{\gamma} +\frac{E^\varphi
  K_\varphi}{2\sqrt{E^x}\gamma}.$ The diffeomorphism and
Hamiltonian constraint can be written as\footnote{To derive equation
  (\ref{hamiltonian}) one substitutes $\bar{A}_\varphi=2\gamma
  K_\varphi$, $2 \gamma K_x=A_x+\eta'$ in equation (47) of reference
  \cite{exterior} and adding the scalar field contribution, e.g
  equation (2) of \cite{husaincollapse}. We have absorbed a factor of
  $4\pi$ in Newton's constant.},  
\begin{eqnarray}
 C_x&=& \frac{1}{G}\left[(E^x)' K_x -E^\varphi (K_\varphi)' \right] -P_\phi \phi',\\
H&=&
\frac{1}{G} \left[
-\frac{1}{2} \frac{E^\varphi}{\sqrt{E^x}}
- 2 K_x K_\varphi \sqrt{E^x} 
-\frac{1}{2} \frac{K_\varphi^2  E^\varphi}{\sqrt{E^x}}
+\frac{1}{8} \frac{\left(E^x\right)'}{\sqrt{E^x}E^\varphi}
-\frac{1}{2}
\frac{\sqrt{E^x}\left(E^x\right)'\left(E^\varphi\right)'}
{\left(E^\varphi\right)^2}
+\frac{1}{2} \frac{\sqrt{E^x}\left(E^x\right)''}{E^\varphi}
\right]\nonumber\\
&&+\frac{1}{2} \frac{P_\phi^2}
{\sqrt{E^x}E^\varphi}
+\frac{1}{2}\frac{\left(E^x\right)^\frac{3}{2}
  \left(\phi'\right)^2}
{E^\varphi}.
\label{hamiltonian}
\end{eqnarray}

Recalling that the total Hamiltonian for the system is given by
$  H_T=\int dx \left(N^x C_x+N H\right)$,
one can redefine the shift $N^x_{\rm new} = N^x_{\rm old} +2 N K_\varphi
\sqrt{E^x}/\left(E^x\right)'$,  and the lapse $N_{\rm new}=N_{\rm old}
\left(E^x\right)'/E^\varphi$, one gets a Hamiltonian constraint that
reads, 
\begin{eqnarray}
  H&=&
\frac{1}{G} 
\left[
-\frac{1}{2} \frac{\left(E^x\right)'\left(1+K_\varphi^2\right)}{\sqrt{E^x}}
+\frac{1}{8}
\frac{\left(\left(E^x\right)'\right)^3}{\left(E^\phi\right)^2
\sqrt{E^x}}
-\frac{1}{2}
\frac{\left(\left(E^x\right)'\right)^2\sqrt{E^x}\left(E^\varphi\right)'}
{\left(E^\varphi\right)^3}
+\frac{1}{2} \frac{\left(E^x\right)'\sqrt{E^x}\left(E^x\right)''}
{\left(E^\varphi\right)^2}
-2 K_\varphi \sqrt{E^x} K_\varphi'\right]\nonumber\\
&&+\frac{1}{2} \frac{\left(\left(E^x\right)'\right)^2P_\phi^2}
{\left(E^\varphi\right)^3\sqrt{E^x}}
+\frac{1}{2} \frac{\left(\left(E^x\right)'\right)^2 
\left(E^x\right)^\frac{3}{2}\left(\phi'\right)^2}
{\left(E^\varphi\right)^3}
-2\frac{K_\varphi\sqrt{E^x}\phi' P_\phi }{E^\varphi}.
\end{eqnarray}
The quantity in the square bracket above is a total derivative,
\begin{eqnarray}
  H&=&\frac{1}{G}\left[
\frac{1}{4} \frac{\left(\left(E^x\right)'\right)^2\sqrt{E^x}}
{\left(E^\varphi\right)^2}-\left(K_\varphi^2+1\right)\sqrt{E^x}\right]'
-\frac{2 K_\varphi\sqrt{E^x}\phi'P_\phi }{E^\varphi}
+\frac{1}{2} \frac{\left(E^x\right)' P_\phi^2}
{\sqrt{E^x}\left(E^\varphi\right)^2}
+\frac{1}{2}\frac{\left(E^x\right)'\left(E^x\right)^\frac{3}{2} 
\left(\phi'\right)^2}{\left(E^\varphi\right)^2}.
\end{eqnarray}
This remarkable property is the key element in allowing to define a
local Hamiltonian. Choosing a gauge in which the term involving the
derivative does not depend on the gravitational variables, one is left
with a Hamiltonian that only depends algebraically on the
gravitational variables. As we mentioned, it appears that this is
typical of all theories in $1+1$ dimensions that involve a mass
function. It at least holds spherically symmetric Einstein gravity in
the traditional and new variables and for the
Callan--Giddings--Harvey--Strominger model.

We will completely gauge
fix the theory. The first gauge condition is $\chi_1=0$ with
\begin{equation}
  \chi_1=E^x-x^2,
\end{equation}
In order to preserve the constraint in
time the Lagrange multiplier $N^x$ gets fixed $N^x=0$ 
The diffeomorphism constraint can be solved, determining the
variable $K_x$. The only constraint left is the 
Hamiltonian, which (omitting an overall factor $1/(G (E^\varphi)^2$) becomes,
\begin{eqnarray}
  H&=&
  \left[x\left(\frac{x^2}{\left(E^\varphi\right)^2}-K_\varphi^2-1\right)
\right]'\left(E^\varphi\right)^2-2 G x K_\varphi \phi' P_\phi
E^\varphi+G P_\phi^2+G x^4 \left(\phi'\right)^2. \label{8}
\end{eqnarray}

Our strategy will be to perform a canonical transformation from the
variables $\phi,P_\phi,K_\varphi,E^\varphi$ to a new set of variables
$X,P_X,f,P_f$ such that $X$ is essentially what appears in the square
bracket differentiated. We will later fix the gauge by setting $X$
equal to a given function of $t,x$.  As a consequence $P_X$, the canonical
momentum of $X$, will not appear differentiated in the
constraint. This means that preserving the gauge fixing condition will
lead to an algebraic equation that determines the lapse, and therefore
to a local true Hamiltonian.

To construct the canonical transformation, let us start by 
identifying the variable $X$,
\begin{equation}
-x\left(\frac{x^2}{\left(E^\varphi\right)^2}-K_\varphi^2-1\right)=
X x^2 \phi^2 +2 G M(t).\label{10}
\end{equation}
Recalling that the scalar field has dimensions of inverse length in
$3+1$ dimensions, the factor $x^2$ on the right is chosen so $X$ has
dimensions of length (or time, since we chose $c=\hbar=1$), since it will
later play the role of time. The factor $\phi^2$ is chosen so weak
fields behave well in the gauge fixing (for instance if $\phi=0$ one
has $K_\varphi=0$ and $E^\varphi=x/\sqrt{1-2GM/x}$ in the usual
Schwarzschild gauge). We added a function of time $M(t)$. Later, if
one studies the fields asymptotically one finds that $M$ is a constant
that corresponds to the ADM mass. At the moment it is just a choice in
the definition of $X$.

To complete the canonical transformation we then seek a generating
function, we choose it to be of type I $F_1(\phi,K_\varphi,X,f)$,
so one has that (recalling that
$\left\{K_\varphi(x),E^\varphi(y)\right\}=
G \delta(x-y)$),
\begin{eqnarray}
 {G}\frac{\partial F_1}{\partial K_\varphi} &=&E^\varphi,\label{11}\\
  \frac{\partial F_1}{\partial \phi} &=&P_\phi,\label{12}\\
  \frac{\partial F_1}{\partial f} &=&-P_f,\\
  \frac{\partial F_1}{\partial X} &=&-P_X.\label{14}
\end{eqnarray}

We start from the first equation and note that we can use (\ref{10})
to write $E^\varphi$ in terms of the quantities that the generating
function depends on,
\begin{equation}
  E^\varphi=-\frac{x}{Y},
\end{equation}
where we chose the minus sign of the square root so the Hamiltonian is
positive definite and for brevity we write,
\begin{equation}
  Y = \sqrt{K_\varphi^2+1-\frac{2GM}{x} -x X
      \phi^2}.
\end{equation}

So we can now proceed to integrate (\ref{11}) and choosing the
integration constant to give the simplest form to the generating
function yields,
$  F_1=-\frac{x}{G}
  \log\left(K_\varphi+Y\right)+\phi f$.

With the generating function and (\ref{12}-\ref{14}) we find the
explicit form of the new variables in terms of the old ones,
\begin{eqnarray}
  P_\phi - \frac{x^2 X \phi}{G Y\left(K_\varphi+Y\right)}&=&f,\label{17}\\
  P_f&=&-\phi,\label{18}\\
  P_X &=& -\frac{x^2 \phi^2}{2 GY\left(K_\varphi+Y\right)}.\label{19}
\end{eqnarray}

The last equation will become the Hamiltonian constraint when we 
rewrite the right hand side entirely in terms of the new variables.
Rewriting $\phi$ is immediate. To obtain $K_\varphi$ we solve
(\ref{8}) rewritten in terms of the new variables, i.e.,
\begin{equation}
  H=-\frac{\left(x^2\phi^2 X\right)'}{Y^2}+ \frac{2 G x^2 K_\varphi
    \phi' P_\phi }{Y}+G P_\phi^2 +G x^4 \left(\phi'\right)^2,
\end{equation}
and $P_\phi$ and $\phi$ given by (\ref{17}) and (\ref{18})
respectively, so we have 
\begin{equation}
  K_\varphi=\frac{x U^2-x+2 M G +X P_f^2 x^2}{2 x U},\label{kvarphi}
\end{equation}
with 
\begin{eqnarray}
  U&=&\frac{1}{\left(f+x^2 P_f'\right) \sqrt{x G}}
\left(
\left(P_f'\right)^2 x^4 G Z
+2 x^5 P_f' X P_f+2 \left(X P_f^2 x^2\right)' x^3
+ 2 X x^3 P_f f
+
f^2 G Z+2 x V^\frac{1}{2}
\right)^\frac{1}{2},
\end{eqnarray}
and
\begin{eqnarray}
  V&=& x\left( \left(P_f'\right)^2 x G Z +
\left(X P_f^2 x^2\right)'\right)\left(f^2 G Z
+2 x^3 X P_f \left(x^2 P_f'  + f \right)+\left(X P_f^2
  x^2\right)'x^3 \right)\\
 Z&=& x^2 X P_f^2 +2 G M -x.
\end{eqnarray}

We now consider (\ref{19}) written entirely in terms of the new
variables,
\begin{equation}
  P_X +\frac{x^3 P_f^2}{2
    GY(f,P_f)\left(K_\varphi(f,P_f)+Y(f,P_f)\right)}=0.
\end{equation}
This expression is the Hamiltonian constraint that is now easy to 
deparameterize. The total Hamiltonian is given by,
$ H_{\rm Total}=\int dx N\left(P_X+{\cal H}_{\rm True}\right)$,
where we recognize the true Hamiltonian density,
\begin{equation}
{\cal H}_{\rm True}  =
\frac{x^3 P_f^2}{2
    GY(f,P_f)\left(K_\varphi(f,P_f)+Y(f,P_f)\right)}.
\end{equation}
To prove that indeed this expression is the true Hamiltonian 
density, we proceed to completely fix the gauge. We choose
$\chi_2=-X+g(x)+t=0$. The preservation in time of this condition,
$\frac{\partial \chi_2}{\partial t} + \left\{\chi_2,H_{\rm Total}\right\}=0
$
 implies that the lapse $N=1$. The system is now totally described in
 terms of the matter field variables $f,P_f$, since $X$ is fixed by
 the gauge fixing and $P_X$ is given by minus the true Hamiltonian. 
If we now consider the time evolution of the remaining variables,
\begin{eqnarray}
  \dot{f}=\left\{f,H_{\rm Total}\right\}=\left\{f,H_{\rm
      True}\right\},\\
  \dot{P_f}=\left\{P_f,H_{\rm Total}\right\}=\left\{P_f,H_{\rm True}\right\},
\end{eqnarray}
showing that the true Hamiltonian indeed generates the evolution.

The expression for $K_\varphi$ (\ref{kvarphi}) contains a series of
square roots. This reflects the fact that the construction will not
work for generic initial data, as one expects in gauge fixed
treatments. In order to analyze under which conditions the
construction works, we  study the situation of weak
fields, so we will assume $f=O(\epsilon)$ and $P_f=O(\epsilon)$ with
$\epsilon\ll 1$ and we will keep only leading terms in $\epsilon$ in
all equations. We will also assume that $M \sqrt{G}\gg 1$ (we are
using units where $\hbar=c=1$). In order to simplify expressions we
will also assume $g(x)=c x$ with $c$ a positive constant. The
expression for $U$ becomes,
\begin{equation}
  U = \frac{1}{f+x^2 P_f}\sqrt{\frac{2 r}{G}}
\left(\left(x^3 c P_f^2\right)'+c x f P_f +x^3 P_f' c P_f +
\left[\left(x^3 c P_f^2\right)'\left(2 c x f P_f +\left(x^3 c
      P_f^2\right)'
+2 x^3 P_f' c P_f\right)\right]^\frac{1}{2}\right)^\frac{1}{2}.
\end{equation}
Sufficient conditions for the existence of the square roots are,

\begin{eqnarray}
\left(x^3 c P_f^2\right)' &=&w(x),\\
\left(2 c x f P_f +2 x^3 P_f' c P_f\right)&=&v(x),
\end{eqnarray}
with   $w(x)$ and $v(x)$ positive functions. Solving the differential
equations we get,
\begin{eqnarray}
  P_f &=& \frac{\sqrt{x \int^x w(x') dx' }}{x^2},\\
  f   &=& \frac{3 x \int^x w(x') dx'-x^2 w(x)}{2x^3 P_f(x)}+\frac{v(x)}{x}.
\end{eqnarray}
So we see that indeed one can specify initial data in the gauge
we chose. 

We have therefore presented for the first time a local Hamiltonian for
a scalar field coupled to gravity in spherical symmetry, a problem
that was unclear had a solution. The technique appears applicable in
other $1+1$ dimensional situations where there exists a mass
function. The result has a counterpart in path integral treatments,
where authors were able to integrate out the gravitational variables
\cite{grumiller}. This includes the
Callan--Giddings--Horowitz--Strominger model, which has received
renewed attention recently \cite{cghs2} and is one of the best
understood models of black hole evaporation. In further work we will
discuss the boundary treatment in these coordinates and will show the
evolution of collapsing scalar field pulses numerically.  The
resulting unconstrained system can be useful for quantization in
situations involving gravitational collapse and black hole
evaporation.

We wish to thank Daniel Grumiller, Viqar Husain, Gabor Kunstatter and
especially Bill Unruh for comments and a referee for corrections.
This work was supported in part by grant NSF-PHY-0968871, funds of the
Hearne Institute for Theoretical Physics, CCT-LSU, Pedeciba and ANII
PDT63/076. This publication was made possible through the support of a
grant from the John Templeton Foundation. The opinions expressed in
this publication are those of the author(s) and do not necessarily
reflect the views of the John Templeton Foundation.

\end{document}